\documentclass[12pt]{article}
\RequirePackage[colorlinks,citecolor=blue,urlcolor=blue]{hyperref}
\usepackage{amsmath}
\usepackage{graphicx}
\usepackage{url} 

\newcommand{\blind}{0}

\input{hzz_app.sty}
\begin{document}

\def\spacingset#1{\renewcommand{\baselinestretch}%
{#1}\small\normalsize} \spacingset{1}


\if0\blind
{
  \title{\bf Accelerating Bayesian inference of dependency between complex biological traits}
  \author{Zhenyu Zhang$^{1}$, \\
  	Akihiko Nishimura$^{2}$, \\
  	N\'{i}dia S. Trov\~{a}o$^{3}$,\\
  	Joshua L. Cherry$^{3,4}$,\\
  	Andrew J.~Holbrook$^1$, \\
  	Xiang Ji$^{5}$, \\
  	Philippe Lemey$^{6}$, \\
  	and Marc A.~Suchard$^{1,7,8}$}  \maketitle
\noindent\it $^1$Department of Biostatistics,  University of California Los Angeles \\
\it $ ^2 $Department of Biostatistics, Johns Hopkins University \\
\it $ ^3 $Division of International Epidemiology and Population Studies, Fogarty International Center, National Institutes of Health \\
\it $ ^4 $National Center for Biotechnology Information, National Library of Medicine, National Institutes of Health \\
\it $ ^5 $Department of Mathematics, Tulane University\\
\it $ ^6 $Department of Microbiology, Immunology and Transplantation, Rega Institute, KU Leuven\\
\it $^7$Department of Biomathematics, \it $^8$Department of Human Genetics, Universtiy of California Los Angeles \\
} \fi
\if1\blind
{
  \bigskip
  \bigskip
  \bigskip
  \begin{center}
    {\LARGE\bf Title}
\end{center}
  \medskip
} \fi

\clearpage
\begin{abstract}
Inferring dependencies between complex biological traits while accounting for evolutionary relationships between specimens is of great scientific interest yet remains infeasible when trait and specimen counts grow large.
The state-of-the-art approach uses a phylogenetic multivariate probit model to accommodate binary and continuous traits via a latent variable framework, and utilizes an efficient bouncy particle sampler (BPS) to tackle the computational bottleneck --- integrating many latent variables from a high-dimensional truncated normal distribution.
This approach breaks down as the number of specimens grows and fails to reliably characterize conditional dependencies between traits.
Here, we propose an inference pipeline for phylogenetic probit models that greatly outperforms BPS. 
The novelty lies in 1) a combination of the recent Zigzag Hamiltonian Monte Carlo (Zigzag-HMC) with linear-time gradient evaluations and 2) a joint sampling scheme for highly correlated latent variables and correlation matrix elements.
In an application exploring HIV-1 evolution from 535 viruses, the inference requires joint sampling from an 11,235-dimensional truncated normal and a 24-dimensional covariance matrix.
Our method yields a 5-fold speedup compared to BPS and makes it possible to learn partial correlations between candidate viral mutations and virulence. 
Computational speedup now enables us to tackle even larger problems: we study the evolution of influenza H1N1 glycosylations on around 900 viruses.
For broader applicability, we extend the phylogenetic probit model to incorporate categorical traits, and demonstrate its use to study \textit{Aquilegia} flower and pollinator co-evolution.
\end{abstract}

\noindent%
\vfill

\newpage
\spacingset{1.5} 
\section{Introduction}
An essential goal in evolutionary biology is to understand the associations between traits observed within biological samples, or \emph{taxa}, ranging from plants and animals to microorganisms and pathogens such as human immunodeficiency virus (HIV) and influenza.
This task is difficult because taxa are implicitly correlated through their shared evolutionary history often described with a reconstructed phylogenetic tree.
Here, tree tips correspond to the taxa themselves, and internal nodes are their unobserved ancestors.
Inferring across-trait covariation requires a highly structured model that can explicitly describe the tree structure and adjust for across-taxa covariation.
Phylogenetic models do exactly this but are computationally challenging because one must integrate out unobserved ancestor traits while accounting for uncertainties arising from tree estimation.
The computational burden increases when taxon and trait counts grow large and becomes worse when traits include continuous and discrete quantities.
\citet{zhang2021large} show that their phylogenetic multivariate probit model provides a promising tool to learn correlations among complex traits at scale when combined with an efficient inference scheme that achieves order-of-magnitudes efficiency gains over the previous best approach \citep{Cybis2015}.
\citet{zhang2021large} demonstrate their method on a data set with $ \nTaxa = 535$ HIV viruses and $ \nTraits=24$ traits that requires sampling from a truncated normal distribution with more than 11,000 dimensions.
In this work, we significantly advance performance compared to \citep{zhang2021large} and solve more challenging problems including the (a) inference of across-trait partial correlations that present clues for potential causal pathways and (b) integration of complex traits with categorical outcomes.

To jointly model complex traits, the phylogenetic probit model assumes discrete traits arise from continuously valued latent variables that follow a Brownian diffusion along the tree \citep{felsenstein1985phylogenies, Cybis2015, zhang2021large}.
Assuming latent processes is a common strategy for modeling complex data and it finds uses across various fields \citep{fedorov2012optimal, schliep2013multilevel, irvine2016extending, pourmohamad2016multivariate, clark2017generalized}.
For $ \nTaxa $ taxa and
$ \nTraits $ continuous or binary traits, Bayesian inference for the phylogenetic probit model involves repeatedly sampling latent variables from their conditional posterior, an $(\nTaxa \times \nTraits)$-dimensional truncated normal distribution.
For this task, \citet{zhang2021large} develop a bouncy particle sampler (BPS) \citep{bouchard2018bouncy} augmented with an efficient dynamic programming approach that speeds up the most expensive step in the BPS implementation.
However, BPS suffers from a major limitation --- it does not allow joint sampling of the latent variables $ \latentData $ and the trait correlation $ \traitCorr $.
\citet{zhang2021large} use a separate Hamiltonian Monte Carlo sampler \citep[HMC]{hmcneal}  to infer $ \traitCorr $ and update the two sets of parameters alternately within a random-scan Gibbs scheme \citep{liu1995covariance}.
Since $\latentData$ and $\traitCorr$ are highly correlated by model assumption, the Gibbs scheme hurts efficiency.

Our solution utilizes a state-of-the-art Markov chain Monte Carlo (MCMC) method called Zigzag-HMC \citep{nishimura2020discontinuous}.
Zigzag-HMC can take advantage of the same $\order{\nTaxa}$ gradient evaluation strategy advanced by \citet{zhang2021large}, yet allows a joint update of $\latentData$ and $\traitCorr$ through differential operator splitting \citep{strang1968construction, nishimura2020discontinuous} which generalizes the previously proposed split HMC framework based on Hamiltonian splitting \citep{hmcneal, shahbaba2014split}.   
The joint sampling scheme greatly improves the mixing of elements in $ \traitCorr $ and thus provides reliable estimates of across-trait partial correlations that describe the conditional dependence between any two traits, free of confounding from other traits in the model.
As seen in our applications, these conditional dependencies provide insights into potential causal pathways driven by real biological processes.

We apply our methodology to three real-world examples.
First, we re-evaluate the HIV evolution application in \citet{zhang2021large} and identify HIV-1 \textit{gag}
immune-escape mutations linked with virulence through strong conditional dependence relationships.
Our findings closely match with the experimental literature and indicate a general pattern in the immune escape mechanism of HIV.
Second, we examine the influenza H1N1 glycosylation pattern across different hosts and detect strong conditional dependencies between glycosylation sites closely related to host switching.
Finally, we investigate how floral traits of \textit{Aquilegia} flower attract different pollinators, for which we generalize the phylogenetic probit model to accommodate a categorical pollinator trait.
\section{Methods}
\subsection{Complex trait evolution}
We describe biological trait evolution with the phylogenetic multivariate probit model following \citet{zhang2021large} and extend it to categorical traits as in \citet{Cybis2015}.
Consider $ \nTaxa $ taxa on a phylogenetic tree $ \phylogeny = (\nodeSet, \branchSet) $ that is a directed, bifurcating acyclic graph.
We either know the tree \emph{a priori} or infer it from a molecular sequence alignment $ \sequenceData $ \citep{beast2018}.
The node set $ \nodeSet $ of size $ 2\nTaxa -1 $ contains $ \nTaxa $ tip nodes, $ \nTaxa - 2 $ internal nodes and one root node.
The branch lengths $ \branchSet = \left( \branchLength{1}, \dots,   \branchLength{2\nTaxa - 2}\right) $ denote the child-parent distance in real time.
We observe $\nTraits$ traits of complex for each taxon.
The trait data $ \observedResponse = \observedResponseMatrix= \left(\observedCont, \observedDiscrete\right)$ partition as $ \observedCont $, an $ \nTaxa \times \nTraitsCont $ matrix of continuous traits and $ \observedDiscrete $,  an $ \nTaxa \times \nTraitsDiscrete $ matrix of discrete ones.
For each node $ i $ in $ \phylogeny $, we assume a $ \dimLatent$-dimensional latent variable $ \latentData_\nodeIndexOne \in \realNumbers^{\dimLatent}$, $i = 1, \dots, 2\nTaxa - 1$, where $\dimLatent = \nTraitsCont + \sum_{j=1}^{\nTraitsDiscrete} \left(\classm{j} - 1 \right) $ and $ \classm{j} $ is the number of classes for the $j$th discrete trait.
To relate latent variables to observed discrete traits, we assume a threshold model for binary traits and a choice model for traits with more than two classes. 
For a categorical trait $ \observedResponseElement $, the possible classes are $ \{\classc{1}, \dots, \classc{\classm{j}}\}$ with the reference class being $ \classc{1}$.
Multiple latent variables $ \latentDataElement{i,j'} ,\dots ,\latentDataElement{i,j'+\classm{j} - 2} $ decide the value of $ \observedResponseElement $.
We summarize the mapping from $ \latentData $ to $ \observedResponse $ as
\begin{equation}\label{eq:thresholdFunc}
\observedResponseElement =
\begin{cases}
\latentDataMatrixElement, & \text{if } \observedResponseElement \text{ is continuous}, \\
\sign(\latentDataElement{ij}), & \text{if } \observedResponseElement \text{ is binary},\\
\classc{1}, & \text{if } \observedResponseElement \text{ is categorical and } M = 0, \\
\classc{\cm}, & \text{if } \observedResponseElement \text{ is categorical}, \cm > 1, \text{and } M = \latentDataElement{i,j' + \cm -2} > 0,
\end{cases}
\end{equation}
where $M =  \max( \latentDataElement{i,j'}, \dots, \latentDataElement{i,j'+\classm{j} - 2}) $ and $ \sign(\latentDataMatrixElement)$ returns the value 1 on positive values and -1 on negative values.
This data augmentation strategy is a common choice to model categorical data \citep{albert1993bayesian}.
As a side note, for continuous $\observedResponseElement$ the corresponding $\latentDataMatrixElement$ is observed, and so $ \latentData_\nodeIndexOne $ is actually a partially latent vector.
Since in our applications only a small fraction of $\observedResponseElement$ is continuous, we omit ``partial" to ease the notation. 

The latent variables follow a multivariate Brownian diffusion process along $ \phylogeny $ such that $ \latentData_{\nodeIndexOne}$ distributes as a multivariate normal (MVN)
\begin{equation}
\latentData_{\nodeIndexOne}  \sim \normalDistribution{\latentData_{\text{pa}\left(\nodeIndexOne\right)}}{\branchLength{\nodeIndexOne}\traitCovariance}, i = 1, ..., 2\nTaxa - 2,
\end{equation}
where $\latentData_{\text{pa}\left(\nodeIndexOne\right)}$ is the parent node value and the $ \dimLatent \times \dimLatent $ covariance matrix $ \traitCovariance $ describes the across-trait association.
The intuition behind $\branchLength{\nodeIndexOne}\traitCovariance$ is that the further away a child node is from its parent node (larger $\branchLength{\nodeIndexOne}$), the bigger difference between their node values.
Assuming a conjugate root prior $ \latentData_{2\nTaxa - 1} \sim \normalDistribution{\rootpriorMean}{\rootpriorSamplesize \inverse \traitCovariance}$ with prior mean $\rootpriorMean$ and prior sample size $\rootpriorSamplesize$, we can analytically integrate out latent variables on all internal nodes.
Marginally, then, the $ \nTaxa \times \dimLatent $ tip latent variables $ \latentData $ have the matrix normal (MTN) distribution
\begin{equation}\label{eq:matrixNormalDistri}
\latentData \sim \matrixNormal{\nTaxa}{\dimLatent}{\tipMeanMatrix}{\phylogenyVariance}{\traitCovariance},
\end{equation}
where $ \tipMeanMatrix = \left(\rootpriorMean,\dots,\rootpriorMean\right)\transpose $ is an $ \nTaxa \times \dimLatent $ mean matrix and the across-taxa covariance matrix  $ \phylogenyVariance$ equals $\diffusionVariance +  \rootpriorSamplesize \inverse \oneMatrix$ \citep{pybus2012}. The tree $ \phylogeny $ determines the diffusion matrix $ \diffusionVariance $ and $ \rootpriorSamplesize \inverse \oneMatrix $ comes from the integrated-out tree root prior, where $ \oneMatrix $ is an all-one $ \nTaxa \times \nTaxa $ matrix.
The augmented likelihood of $ \latentData $ and $ \observedResponse $ factorizes as
\begin{equation}\label{eq:3}
\cDensity{\observedResponse, \latentData}{\phylogenyVariance, \traitCovariance,\rootpriorMean, \rootpriorSamplesize}  = \cDensity{\observedResponse}{\latentData}  \cDensity{\latentData}{\phylogenyVariance,\traitCovariance,\rootpriorMean, \rootpriorSamplesize},
\end{equation}
where $ \cDensity{\observedResponse}{\latentData} = 1$ if $ \latentData $ are consistent with $ \observedResponse $ according to Equation \eqref{eq:thresholdFunc} and 0 otherwise.
Following \citet{zhang2021large}, we decompose $ \traitCovariance$ as $\traitDiag \traitCorr  \traitDiag$ such that $ \traitCorr $ is the  $ \dimLatent \times \dimLatent $ correlation matrix and $  \traitDiag$ is a diagonal matrix with marginal standard deviations.
Importantly, since discrete traits only inform the sign or ordering of their underlying latent variables, certain elements of $\traitDiag$ must be set as a fixed value to ensure that the model is parameter-identifiable.
\citet{zhang2021large} demonstrate the necessity of this $\traitDiag \traitCorr  \traitDiag$ decomposition, which also allows a non-informative prior \citep[LKJ]{lkj2009} on $\traitCorr$.
For goodness-of-fit of the phylogenetic probit model we refer interested readers to \citet{zhang2021large} where the explicit tree modeling leads to a significantly better fit. 

\subsection{A novel inference scheme}\label{sec:splitHMC}
We sample from the joint posterior to learn the across-trait correlation $ \traitCorr$
\begin{equation}
\begin{aligned}
\cDensity{\traitCorr,\traitDiag, \latentData, \phylogeny}{\observedResponse,\sequenceData} &\propto \cDensity{\observedResponse}{\latentData} \:\times\: \cDensity{\latentData}{\traitCorr,\traitDiag, \phylogeny} \:\times\: \\& \quad  \quad  \density{\traitCorr, \traitDiag} \:\times\: \cDensity{\sequenceData}{\phylogeny} \:\times\: \density{\phylogeny},
\end{aligned}
\end{equation}
where we drop the dependence on hyper-parameters $(\phylogenyVariance,\rootpriorMean, \rootpriorSamplesize)$ to ease notation.
We then specify the priors $ \density{\traitCorr, \traitDiag} $ and $ \density{\phylogeny} $ as in \citet{zhang2021large}.
Assuming $ \density{\traitCorr, \traitDiag} = \density{\traitCorr}\density{\traitDiag}$ and an LKJ prior on $ \traitCorr $, we set independent log normal priors on $ \traitDiag $ diagonals that correspond to discrete traits, and assume a typical coalescent tree prior on $\phylogeny$ \citep{kingman1982coalescent}.
\citet{zhang2021large} use a random-scan Gibbs \citep{liu1995covariance} scheme to alternately update  $\latentData$, $\{\traitCorr, \traitDiag\}$ and $\phylogeny$ from their full conditionals \citep{beast2018}.
They sample $\latentData$ from an $\nTaxa \dimLatent$-dimensional truncated normal distribution with BPS and deploy the standard HMC based on Gaussian momentum \citep{hoffman2014nuts} to update $\{\traitCorr, \traitDiag\}$.
Instead, we simulate the joint Hamiltonian dynamics on $ \{\latentData, \traitCorr, \traitDiag\} $ by combining novel Hamiltonian zigzag dynamics on $ \latentData $ \citep{nishimura2021hamiltonian} and traditional Hamiltonian dynamics on $\{\traitCorr, \traitDiag\}$.
This strategy enables an efficient joint update of the two highly-correlated sets of parameters.
We first describe how Zigzag-HMC samples $ \latentData $ from a truncated normal and then detail the joint update of $ \{\latentData, \traitCorr, \traitDiag\} $.

\subsubsection{Zigzag-HMC for truncated multivariate normals}\label{sec:HZZonGaussian}
We outline the main ideas behind HMC \citep{hmcneal} before describing Zigzag-HMC as a version of HMC based on \textit{Hamiltonian zigzag dynamics} \citep{nishimura2020discontinuous, nishimura2021hamiltonian}.
In order to sample a $ \paramDimension $-dimensional parameter $ \bposition = \left(\position_1, \dots, \position_\paramDimension\right)$ from the target distribution $ \pi(\bposition) $, HMC introduces an auxiliary \textit{momentum} variable $ \bmomentum = \left(\momentum_1, \dots, \momentum_\paramDimension\right)\in \realNumbers^\paramDimension$ and samples from the product density $\pi(\bposition, \bmomentum) = \pi(\bposition) \pi(\bmomentum)$ by numerically discretizing the Hamiltonian dynamics
\begin{equation}
\label{eq:hamilton}
\frac{\diff \bposition}{\diff t}
= \nabla \kinetic(\bmomentum), \quad
\frac{\diff \bmomentum}{\diff t}
= - \nabla \potential(\bposition),
\end{equation}
where $\potential(\bposition)=- \log \pi(\bposition)$ and $\kinetic(\bmomentum) = - \log \pi(\bmomentum)$ are the potential and kinetic energy.
In each HMC iteration, we first draw $\bmomentum$ from its marginal distribution $ \momentumDistribution{\bmomentum} \sim \mathcal{N}(\bm{0}, \M) $, a standard Gaussian and then approximate \eqref{eq:hamilton} from time $t = 0$ to $t = \tau$ by $L = \lfloor \tau / \stepsize \rfloor$ steps of the \textit{leapfrog} update with stepsize $\epsilon$ \citep{leimkuhler2004simulating}:
\begin{equation}\label{eq:leapfrog}
\bmomentum  \gets \bmomentum + \frac{\stepsize}{2} \nabla_{\bposition} \log \pi(\bposition), \quad
\bposition \gets \bposition + \stepsize  \bmomentum, \quad
\bmomentum \gets \bmomentum + \frac{\stepsize}{2} \nabla_{\bposition} \log \pi(\bposition).
\end{equation}
The end state is a valid \textit{Metropolis} proposal that one accepts or rejects according to the standard acceptance probability formula \citep{metropolis53, hastings1970monte}.

Zigzag-HMC differs from standard HMC insofar as it posits a Laplace momentum $ \momentumDistribution{\bmomentum} \propto \prod_{i} \exp\left(-|\momentum_i|\right), i = 1, \dots, \paramDimension$.
The Hamiltonian differential equations now become
\begin{equation}
\label{eq:hzz_equation}
\frac{{\rm d} \bposition}{{\rm d} t}
=   \sign\left(\bmomentum\right), \quad
\frac{{\rm d} \bmomentum}{{\rm d} t} = - \nabla \potential(\bposition),
\end{equation}
and the velocity $\bvelocity := \diff \bposition / \diff t \in \{\pm 1\}^\nParam$ depends only on the sign of $\bmomentum$ and thus remains constant until one of $ \momentum_i $'s undergoes a sign change (an ``event").
To understand how the Hamiltonian zigzag dynamics \eqref{eq:hzz_equation} evolve over time, one must investigate when such events happen.
Before moving to the truncated MVN, we first review the event time calculation for a general $  \pi(\bposition) $ following \citet{nishimura2021hamiltonian}.
Let $ \eventTk{k}$ be the $ k $th event time and $\left( \bposition\left(\eventTk{0}\right), \bv\left(\eventTk{0}\right), \bmomentum\left(\eventTk{0}\right)\right)$ is the initial state at time $ \eventTk{0} $.
Between $ \eventTk{k} $ and $ \eventTk{k+1} $, $ \bposition $ follows a piecewise linear path and the dynamics evolve as
\begin{equation}
\label{eq:position_evolution}
\bposition\bigl(\eventTk{k} + t\bigr) = \bposition\bigl(\eventTk{k}\bigr) + t\bvelocity\bigl(\eventTk{k}\bigr), \quad
\bvelocity\bigl(\eventTk{k} + t\bigr) = \bvelocity\bigl(\eventTk{k}\bigr), \quad t \in \bigl[0, \eventTk{k+1} - \eventTk{k}\bigr),
\end{equation}
and
\begin{equation}
\label{eq:momentum_evolution}
\momentum_i\bigl(\eventTk{k} + t\bigr) =  \momentum_i\bigl(\eventTk{k}\bigr) - \int_{0}^{t} \partial_i \potential \left[\bposition\bigl(\eventTk{k}\bigr) + s\bvelocity\bigl(\eventTk{k}\bigr)\right]ds \quad \text{for } i = 1,\dots, \paramDimension.
\end{equation}
Therefore we can derive the $ (k+1) $th event time
\begin{equation}
\label{eq:coord_event_time}
\eventTk{k+1} = \eventTk{k} + \min_{i}\eventTimeI,  \quad
\eventTimeI = \min_{t>0}\biggl\{
\momentum_i\bigl(\eventTk{k}\bigr) = \int_{0}^{t} \partial_i \potential \bigl[\bposition(\eventTk{k}) + s\bvelocity(\eventTk{k})\bigr]ds
\biggr\},
\end{equation}
and the dimension causing this event is $\istar = \argmin_i \eventTimeI$.
At the moment of $ \eventTk{k+1} $, the $ \istar $th velocity component flips its sign
\begin{equation}
\velocity_{\istar}\bigl(\eventTk{k+1}\bigr) = -\velocity_{\istar}\bigl(\eventTk{k}\bigr), \quad \velocity_{j}\bigl(\eventTk{k+1}\bigr) = \velocity_{j}\bigl(\eventTk{k}\bigr) \text{ for } j\neq \istar.
\end{equation}
Then the dynamics continue for the next interval $ \bigl[\eventTk{k+1}, \eventTk{k+2}\bigr) $.

We now consider simulating the Hamiltonian zigzag dynamics for a $d$-dimensional truncated MVN defined as
\begin{equation}
\label{eq:bps_targetDistribution_tMVN}
\bposition \sim \normalDistribution{\bmu}{\grandVariance}\text{ subject to } \bposition \in \{\map(\bposition) = \observedResponseVec \},
\end{equation}
where $ \observedResponseVec \in \realNumbers^\nTraits$ is the complex data, $ \map(\cdot) $ is the mapping from latent variables $ \bposition $ to $ \observedResponseVec $ as in Equation \eqref{eq:thresholdFunc}, $ \bposition \in \realNumbers^ \paramDimension $ and $ \paramDimension \geq \nTraits $.
In this setting, we have $\nabla \potential(\bposition) = \grandVariance\inverse \bposition$ whenever $\bposition \in \{\map(\bposition) = \observedResponseVec \}$.
Importantly, this structure allows us to simulate the Hamiltonian zigzag dynamics exactly and efficiently \citep{nishimura2021hamiltonian}.
We handle the constraint $\map(\bposition) = \observedResponseVec$ with a technique from \citet{hmcneal} where the constraint boundaries embody ``hard walls" that the Hamiltonian zigzag dynamics ``bounce" against upon impact.
To distinguish different types of events, we define \textit{gradient events} arising from solutions of Equation \eqref{eq:coord_event_time}, \textit{binary events} arising from hitting binary data boundaries and \textit{categorical events} arising from hitting categorical data boundaries.

We first consider how to find the gradient event time.
Starting from a state $ \left(\bposition,  \bvelocity, \bmomentum \right) $, by plugging in $\nabla \potential(\bposition) = \grandVariance\inverse \bposition$ to Equation~\eqref{eq:coord_event_time}, we can calculate the gradient event time $ \gradientTime $ by first solving $ \paramDimension $ quadratic equations
\begin{equation}
\label{eq:zigzag_event_time}
\bmomentum = t \grandVariance\inverse (\bposition - \bmu) + \frac{t^2}{2} \grandVariance\inverse \bv,
\end{equation}
and then taking the minimum among all positive roots of Equation \eqref{eq:zigzag_event_time}.
When $ \pi(\bposition) $ is a truncated MVN arising from the phylogenetic probit model, we exploit the efficient gradient evaluation strategy in \citet{zhang2021large} to obtain $\grandVariance\inverse (\bposition - \bmu)$ and $\grandVariance\inverse \bv$ without the notorious $\order{\paramDimension^3} $ cost to invert $ \grandVariance $.

Next, we focus on the binary and categorical events.
\sloppy{We partition $ \bposition $ into three sets: $ \positionSet{\text{cont}} = \{\position_i:\position_i \text{ is for continuous data} \}$, $ \positionSet{\text{bin}} = \{\position_i: \position_i \text{ is for binary data} \} $, and $\positionSet{\text{cat}} = \{\position_i:\position_i \text{ is for categorical data} \} $.}
Since latent variables in $ \positionSet{\text{cont}}  $ are fixed, we ``mask" them out following \citet{zhang2021large}.
Starting from a state $ \left(\bposition, \bvelocity, \bmomentum\right) $, a binary event happens at time $ \reflectTime $ when the trajectory first reaches a binary boundary at dimension $ i_\textrm{b} $
\begin{equation}
\label{eq:find_bd_time}
\reflectTime = \left| x_{i_\textrm{b}} / v_{i_\textrm{b}} \right|, \
{i_\textrm{b}} = \textstyle \argmin_{\, i \in I_{\text{bin}}} \left| x_i / v_i \right|
\, \text{ for } \,
I_{\text{bin}} = \{i : x_i v_i < 0 \text{ and } \position_i \in \positionSet{\text{bin}} \}.
\end{equation}
Here, we only need to check the dimensions satisfying $ x_i v_i < 0 $, i.e., those for which the trajectory is heading towards the boundary.
At time $ \reflectTime $, the trajectory bounces against the binary boundary, and so the $ i_\textrm{b} $th velocity and momentum element both undergo an instantaneous flip $\velocity_{i_\textrm{b}} \gets  -\velocity_{i_\textrm{b}}$, $\momentum_{i_\textrm{b}} \gets  -\momentum_{i_\textrm{b}}$, while other dimensions stay unchanged.

Finally, we turn to categorical events.
Suppose that a categorical trait $y_j = \classc{m} $ belongs to one of $ n $ possible classes, and $ \position_1, \position_2,\dots, \position_{n-1} $ the underlying latent variables.
Equation \eqref{eq:thresholdFunc} specifies the boundary constraints.
If $ \cm = 1 $, the $n-1$ latent variables must be all negative, which poses the same constraint as if they were for $ n-1 $ binary traits, therefore we can solve the event time using Equation \eqref{eq:find_bd_time}.
If $ \cm > 1 $, we must check when and which two dimensions first violate the order constraint $\latentDataElement{\cm-1} = \max( \latentDataElement{1}, \dots, \latentDataElement{n-1}) > 0$.
With the dynamics starting from $ \left(\bposition, \bvelocity, \bmomentum\right) $, the categorical event time $ \cateTime^j $ is given by
\begin{equation}
\label{eq:cateTime}
\begin{aligned}
\cateTime^j
= \left| (x_{\cm-1} - x_{i_\textrm{c}}) / (v_{\cm-1} - v_{i_\textrm{c}}) \right|, &\
{i_\textrm{c}} = \textstyle \argmin_{\, i \in I_{\text{cat}}} \left| (x_{\cm-1} - x_{i}) / (v_{\cm-1} - v_{i}) \right|,
\\ & \text{ for } \,
I_{\text{cat}}= \{i : v_{\cm-1}  < v_i \text{ and } \position_i \in \positionSet{\text{cat}} \},
\end{aligned}
\end{equation}
when $ \latentDataElement{i_\textrm{c} }$ reaches $ \latentDataElement{\cm-1} $ and violates the constraint.
To identify $i_\textrm{c}$ we only need to check dimensions with $v_{\cm-1}  < v_i$ where the distance $ \latentDataElement{\cm-1}-\latentDataElement{i} $ is decreasing.
At $ \cateTime^j $, the two dimensions involved ($ \cm-1 $ and $ i_\textrm{c} $) bounce against each other such that $ \velocity_{\cm-1} \gets  -\velocity_{\cm-1}$, $ \velocity_{ i_\textrm{c} } \gets  -\velocity_{i_\textrm{c} } $, $\momentum_{\cm-1} \gets  -\momentum_{\cm-1} $, $
\momentum_{i_\textrm{c} } \gets  -\momentum_{i_\textrm{c}}$.
Note $\cateTime^j$ is for a single $ y_j $ and we need to consider all categorical data to find the actual categorical event time $ \cateTime = \min_j \cateTime^j $.

We now present the dynamics simulation with all three event types included, starting from a state $ \left(\bposition, \bvelocity, \bmomentum\right) $ with $\bposition \in \{\map(\bposition) = \observedResponseVec \}$:
\begin{enumerate}
	\item Solve $ \gradientTime$,  $\reflectTime$, $\cateTime $ using Equations  \eqref{eq:zigzag_event_time}, \eqref{eq:find_bd_time} and \eqref{eq:cateTime} respectively.
	\item Determine the actual (first) event time $ t = \min\{\gradientTime, \reflectTime, \cateTime\} $ and update $ \bposition $ and $ \bmomentum $ as in Equations \eqref{eq:position_evolution} and \eqref{eq:momentum_evolution} for a duration of $ t$.
	\item Make instantaneous velocity and momentum sign flips according to the rules of the actual event type, then go back to Step 1.
\end{enumerate}
Based on the above discussion, Algorithm~\ref{alg:laplace_hmc} describes one iteration of Zigzag-HMC on truncated MVNs where we simulate the Hamiltonian zigzag dynamic for a pre-specified duration $ \totalTime $.
For a truncated MVN arising from the phylogenetic probit model, we adopt the dynamic programming strategy of \citet{zhang2021large} to speed up the most expensive gradient evaluation step in line \ref{line:expensiveGradient1} and reduce its cost from $\order{\nTaxa^2 \dimLatent +\nTaxa \dimLatent^2}$ to $ \order{\nTaxa\dimLatent^2} $.
In brief, this strategy avoids explicitly inverting $ \phylogenyVariance$ by recursively traversing the tree \citep{pybus2012} to obtain $ \nTaxa $ conditional densities that directly translate to the desired gradient.

\makeatletter
\newcommand{\multiline}[1]{%
	\begin{tabularx}{\dimexpr\linewidth-\ALG@thistlm}[t]{@{}X@{}}
		#1
	\end{tabularx}
}
\makeatother
\begin{algorithm}[htp]
	\setstretch{1.07}
	\caption{Zigzag-HMC for multivariate truncated normal distributions}
	\label{alg:laplace_hmc}
	\begin{algorithmic}[1]
		\Function{HzzTMVN}{$\bposition, \bmomentum,\totalTime$}
		\State $\bv \gets  \sign(\p)$
		\State $\bm{\varphi}_{\bposition} \gets \bPhi(\bposition - \bmu)$ \label{line:expensiveGradient1}
		\State $\remainTime \gets \totalTime$
		\vspace{2mm}
		\While{$\remainTime > 0$}
		\vspace{2mm}
		\LeftComment{find gradient event time $ \gradientTime $}
		\State $ \bm{a} \gets \bm{\varphi}_{\bv} / 2, \bm{b} \gets \bm{\varphi}_{\position}, \bm{c} \gets - \p $
		\State $\gradientTime \gets \min_i \left\{\text{minPositiveRoot}(a_i, b_i, c_i)\right\}$ \Comment{``minPositiveRoot'' defined below}
		
		\vspace{2mm}
		\LeftComment{find binary boundary event time }
		\State $\reflectTime \gets \min_i x_{\nodeIndexOne} /  v_\nodeIndexOne, \text{ for } \nodeIndexOne \text{ with }x_{\nodeIndexOne}  v_\nodeIndexOne < 0 $ and $ \position_i \in \positionSet{\text{bin}} $
		
		\vspace{2mm}
		\LeftComment{find categorical boundary event time, $ n_c $ = number of categorical traits}
		\For{$j = 1,\dots, n_c$}
		\State \multiline{$ \cateTime^j \gets  \min_{\, i} \left| (x_{k-1} - x_{i_\textrm{c}}) / (v_{k-1} - v_i) \right| \text{ for } \nodeIndexOne \text{ with }  v_{k-1}< v_i $ and $ \position_i \in \positionSet{\text{cat}} $}
		\EndFor
		\State $\cateTime \gets \min_j \cateTime^j $
		
		\vspace{2mm}
		\LeftComment{the actual event happens at time $ \bounceTime $}
		\State $\bounceTime  \gets \min \left\{ \gradientTime, \reflectTime, \cateTime, \remainTime \right\} $
		\State $\bposition \gets \bposition + \bounceTime \bm{v}$,
		$\p \gets \p - \bounceTime \bm{\varphi}_{\bposition} - \bounceTime^2 \bm{\varphi}_{\bv} / 2$,
		$\bm{\varphi}_{\bposition} \gets \bm{\varphi}_{\bposition} + \bounceTime \bm{\varphi}_{\bv}$
		\If {a gradient event happens at $ i_g $}
		\State $\velocity_{i_g} \gets -\velocity_{i_g}$
		\ElsIf {a binary boundary event happens at $ i_\textrm{b} $}
		\State $\velocity_{i_\textrm{b}} \gets -\velocity_{i_\textrm{b}}$, $\momentum_{i_\textrm{b}} \gets -\momentum_{i_\textrm{b}}$
		\ElsIf {a categorical boundary event happens at $ i_{\textrm{c1}}, i_{\textrm{c2}}$}
		\State  $\velocity_{i_\textrm{c1}} \gets -\velocity_{i_\textrm{c1}},\velocity_{i_\textrm{c2}} \gets -\velocity_{i_\textrm{c2}},\momentum_{i_\textrm{c1}} \gets -\momentum_{i_\textrm{c1}},\momentum_{i_\textrm{c2}} \gets -\momentum_{i_\textrm{c2}}$
		\EndIf
		\State $\bm{\varphi}_{\bv} \gets \bm{\varphi}_{\bv} +  2 v_{i} \bpsPrecision \basisVector{i}$
		\State $\remainTime \gets \remainTime - \bounceTime$
		\EndWhile \\
		\Return $ \bposition,\bmomentum$
		\EndFunction
	\end{algorithmic}
	\vspace{5mm}
	\algcomment{*{\setstretch{1} $\text{minPositiveRoot}(a_i, b_i, c_i) $ returns the minimal positive root of the equation $ a_i x^2 + b_i x + c = 0 $, or else returns $+\infty$ if no positive root exists.\par}}
\end{algorithm}
\subsubsection{Jointly updating latent variables and across-trait covariance}\label{sec:jointUpdate}
The $ \nTaxa \times \dimLatent $ latent variables and $ \dimLatent \times \dimLatent $ across-trait covariance are highly correlated with each other, so individual Gibbs updates can be inefficient. 
The posterior conditional of $\latentData$ is truncated normal and thus allows for the efficient Hamiltonian zigzag simulation as described in Section~\ref{sec:HZZonGaussian}.
The conditional distribution for covariance components $\traitCorr$ and $\traitDiag$ has no such special structure, so we map them to an unconstrained space and deploy Hamiltonian dynamics based on Gaussian momentum.
We use a standard mapping of $ \traitCorr $ elements to real numbers \citep{stan18} that first transforms $ \traitCorr $ to canonical partial correlations (CPC) that fall in $ [-1,1] $ and then apply the Fisher transformation to map CPC to the real line.
We then construct the joint update of latent variables and covariance via differential operator splitting \citep{strang1968construction, nishimura2020discontinuous} to approximate the joint dynamics of Laplace-Gauss mixed momenta.

We denote the two concatenated sets of parameters $ \latentData $  and $\{\traitCorr$, $\traitDiag\}$ as $ \bposition = \left(\bposition_\textrm{G}, \bposition_\textrm{L}\right) $ with momenta $ \bmomentum = \left(\bmomentum_\textrm{G}, \bmomentum_\textrm{L}\right) $, where indices G  and L refer to Gaussian or Laplace momenta.
The joint sampler updates $ \left(\bposition_G, \bmomentum_G\right) $ first,  then $ \left(\bposition_L, \bmomentum_L\right) $, followed by another update of $ \left(\bposition_G, \bmomentum_G\right) $.
This symmetric splitting ensures that the simulated dynamics is reversible and hence constitute a valid \textit{Metropolis} proposal mechanism \citep{nishimura2020discontinuous}. 
The LG-STEP function in Algorithm \ref{alg:splithmc}  describes the process of simulating the joint dynamics for time duration $2 \epsilon$ via the analytical Hamiltonian zigzag dynamics for  $ \left(\bposition_L, \bmomentum_L\right) $ and the approximate leapfrog dynamics \eqref{eq:leapfrog} for $\left(\bposition_G, \bmomentum_G\right) $.
Because $ \bposition_G $ and $ \bposition_L $ can have very different scales, we incorporate a tuning parameter, the step size ratio $ \rs $, to allow different step sizes for the two dynamics.
To approximate a trajectory of the joint dynamics from $t = 0$ to $t = \tau$, we apply the function LG-STEP $\mstep = \lfloor \tau / 2 \epsilon \rfloor$ times, and accept or reject the end point following the standard acceptance probability formula \citep{metropolis53, hastings1970monte}.
We call this version of HMC based on Laplace-Gauss mixed momenta as \textit{LG-HMC} and describe one iteration of LG-HMC in Algorithm \ref{alg:splithmc} where the inputs include the joint potential function $\potential (\bposition_G, \bposition_L)$. 
We use LG-HMC to update $ \{\latentData, \traitCorr, \traitDiag\} $ as a Metropolis-within-Gibbs step of our random-scan Gibbs scheme.
The overall sampling efficiency strongly depends on $ \mstep $, the step size $ \stepsize $ and the step size ratio $ \rs $, so it is preferable to auto-tune all of them.
Appendix \ref{sec:tuning} provides an empirical method to automatically tune $ \rs $.
We utilize the no-U-turn algorithm to automatically decide the trajectory length $ \mstep $ \citep{hoffman2014nuts} and call the resulting algorithm \textit{LG No-U-Turn Sampler} (LG-NUTS).
We adapt the step size $ \stepsize $ with primal-dual averaging to achieve an optimal acceptance rate \citep{hoffman2014nuts}.
\begin{algorithm}[htp]
	\setstretch{1.2}
	\caption{One LG-HMC iteration}
	\label{alg:splithmc}
	\begin{algorithmic}[1]
		\Function{LG-HMC}{$\bposition_G, \bposition_L , \bmomentum_G, \bmomentum_L , \potential, \mstep, \stepsize, \rs$}
		\LeftComment{Record the initial state}
		\State $ \bposition_G^0 \gets \bposition_G, \bposition_L^0  \gets \bposition_L, \bmomentum_G^0 \gets \bmomentum_G, \bmomentum_L^0 \gets \bmomentum_L$
		\For{$i = 1,\dots, \mstep$}
		\State $ \bposition_G, \bposition_L , \bmomentum_G, \bmomentum_L \gets $ \Call{LG-STEP}{$\bposition_G, \bposition_L , \bmomentum_G, \bmomentum_L , \stepsize, \rs$}
		\EndFor
		\vspace{2mm}
		\LeftComment{Calculate the acceptance probability $\acProb $, where $ \kinetic_G$ and $ \kinetic_L$ denote the kinetic energy based on Gaussian or Laplace momentum and $ \lonenorm{\cdot} $, $ \ltwonorm{\cdot} $ are the $ L^1 $ and $ L^2 $ norm.}
		\State $ \kinetic_G^0 \gets \left(\ltwonorm{\bmomentum_G^0}\right)^2 /2$,  $ \kinetic_L^0 \gets \lonenorm{\bmomentum_L^0}$ 
		\State $ \kinetic_G \gets \left(\ltwonorm{\bmomentum_G}\right)^2 /2$,  $ \kinetic_L \gets \lonenorm{\bmomentum_L}$
		\State $ \acProb \gets \min \{1, \exp \left[\potential(\bposition_G^0, \bposition_L^0) - \potential(\bposition_G, \bposition_L) + \kinetic_G^0 +  \kinetic_L^0 - \kinetic_G - \kinetic_L\right]\} $
		\vspace{2mm}
		\LeftComment{Accept or reject}
		\State $ u \gets$ one draw from uniform$ (0,1) $
		\If {$ u < a$ }
		\State \Return $\bposition_G, \bposition_L , \bmomentum_G, \bmomentum_L$
		\Else
		\State \Return $\bposition_G^0, \bposition_L^0, \bmomentum_G^0, \bmomentum_L^0$
		\EndIf
		\EndFunction
		
		\vspace{2mm}
		\Function{LG-STEP}{$\bposition_G, \bposition_L , \bmomentum_G, \bmomentum_L , \stepsize, \rs$}
		\State $ \bposition_G, \bmomentum_G \gets $ \Call{LeapFrog}{$ \bposition_G, \bmomentum_G , \stepsize $}
		\State $ \bposition_L, \bmomentum_L \gets $ \Call{HzzTMVN}{$ \bposition_G, \bmomentum_G , \rs\stepsize $}
		\State $ \bposition_G, \bmomentum_G \gets $ \Call{LeapFrog}{$ \bposition_G, \bmomentum_G , \stepsize $}
		\State \Return $\bposition_G, \bposition_L , \bmomentum_G, \bmomentum_L$
		\EndFunction
		
		\vspace{2mm}
		\Function{LeapFrog}{$ \bposition_G, \bmomentum_G , \stepsize $}
		\State $ \bmomentum_G \gets   \bmomentum_G + \frac{\stepsize}{2} \nabla_{\bposition_G} \log p(\bposition)$
		\State $ \bposition_G \gets   \bposition_G + \stepsize \bmomentum_G$
		\State $ \bmomentum_G \gets   \bmomentum_G + \frac{\stepsize}{2} \nabla_{\bposition_G} \log p(\bposition)$
		\State \Return $\bposition_G, \bposition_L$
		\EndFunction
	\end{algorithmic}
\end{algorithm}

\section{Results}
We demonstrate the superior efficiency of our joint inference scheme on learning dependency between traits under the phylogenetic probit model, as compared to the state-of-the-art BPS.
To illustrate the broad applicability of our method, we detail three real-world applications and discuss the scientific findings.  
In Section \ref{sec:hiv} we apply our method to the HIV virulence application of \citet{zhang2021large}.
The improved efficiency (shown in Section \ref{sec:effCompare}) allows us to estimate the across-trait partial correlation with adequate effective sample size (ESS) and to reveal the conditional dependence among traits of scientific interest.
We use the same HIV data set to demonstrate that LG-HMC and LG-NUTS outperform BPS (Section \ref{sec:effCompare}), followed by two more LG-NUTS applications on influenza (Section \ref{sec:glyco}) and \textit{Aquilegia} flower (Section \ref{sec:flower}) evolution.
We conclude this section with MCMC convergence criteria and timing results (Section \ref{sec:converge}).

\subsection{HIV immune escape}
\label{sec:hiv}
In the HIV evolution application of \citet{zhang2021large}, a main scientific focus lies on the association between HIV-1 immune escape mutations and virulence, the pathogen's ability to cause disease.
The human leukocyte antigen (HLA) system is predictive of the disease course as it plays an important role in the immune response against HIV-1.
Through its rapid evolution, HIV-1 can acquire mutations that aid in escaping HLA-mediated immune response, but the escape mutations may reduce its fitness and virulence \citep{nomura2013significant,Payne2014}.
\citet{zhang2021large} identify HLA escape mutations associated with virulence while controlling for the unknown evolutionary history of the viruses.
However, \citet{zhang2021large} interpret their results based on the across-trait correlation $ \traitCorr $ which only informs marginal associations that can remain confounded.
Now armed with a more efficient inference method, we are able to focus on the across-trait partial correlation matrix $\traitPcorr = \{\traitPcorrElement_{ij}\}$ that indicates the conditional dependency between two interested traits without confounding from other factors.   
We obtain $ \traitPcorr $ by transforming the inferred $\traitCovariance$ through
\begin{equation}
\traitCovariance \inverse  = \traitPrecision = \{\traitPrecisionElement_{ij}\}, \quad
\traitPcorrElement_{ij} = - \dfrac{\traitPrecisionElement_{ij}}{\sqrt{\traitPrecisionElement_{ii} \traitPrecisionElement_{jj}}}.
\end{equation}

\begin{figure}[htp]
	\centering
	\begin{subfigure}{0.45\textwidth}
		\includegraphics[width=\textwidth]{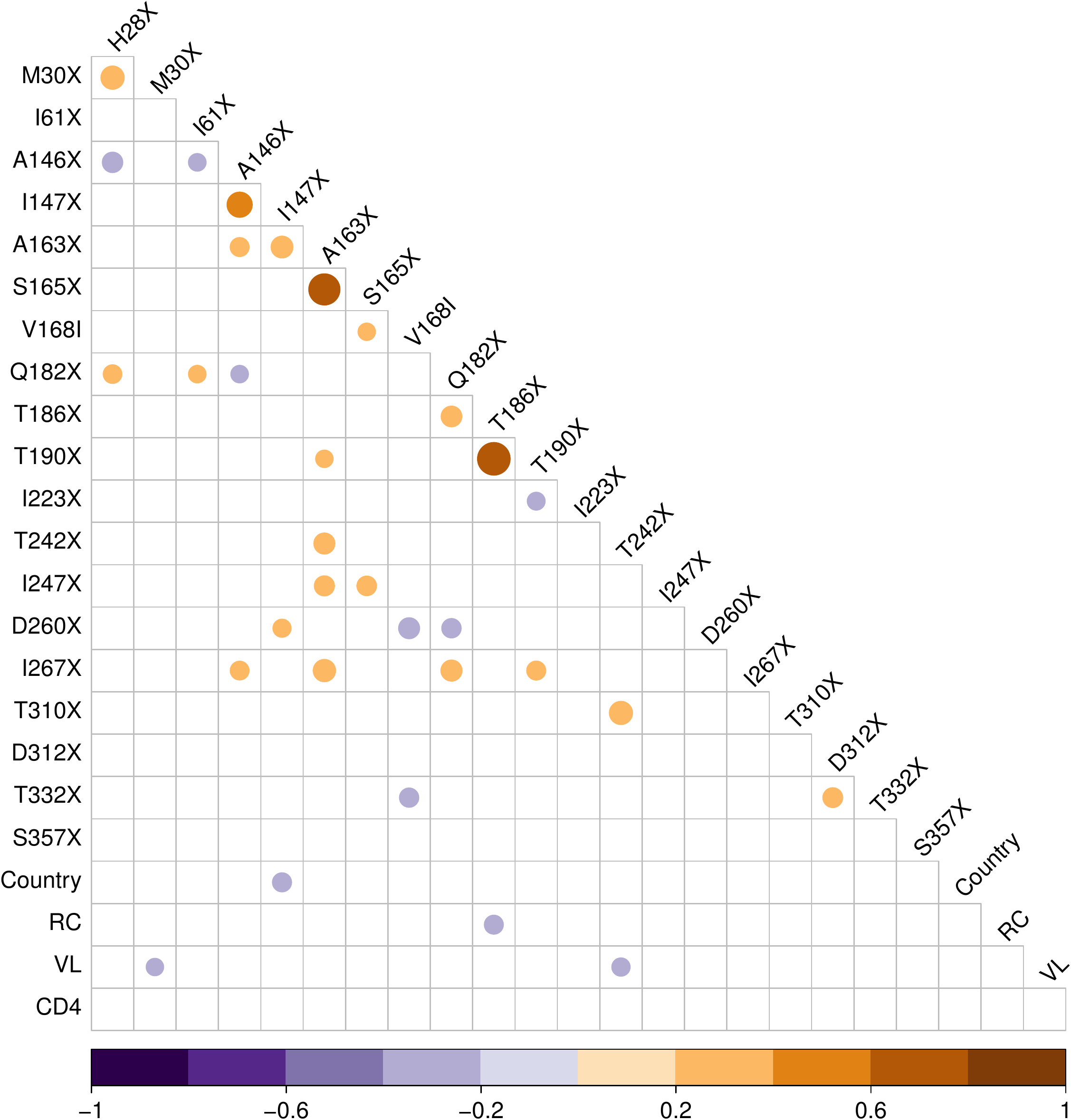}
		\caption{Correlation}
		\label{fig:hiv_corr}
	\end{subfigure}
	\hspace{1em}
	\begin{subfigure}{0.45\textwidth}
		\includegraphics[width=\textwidth]{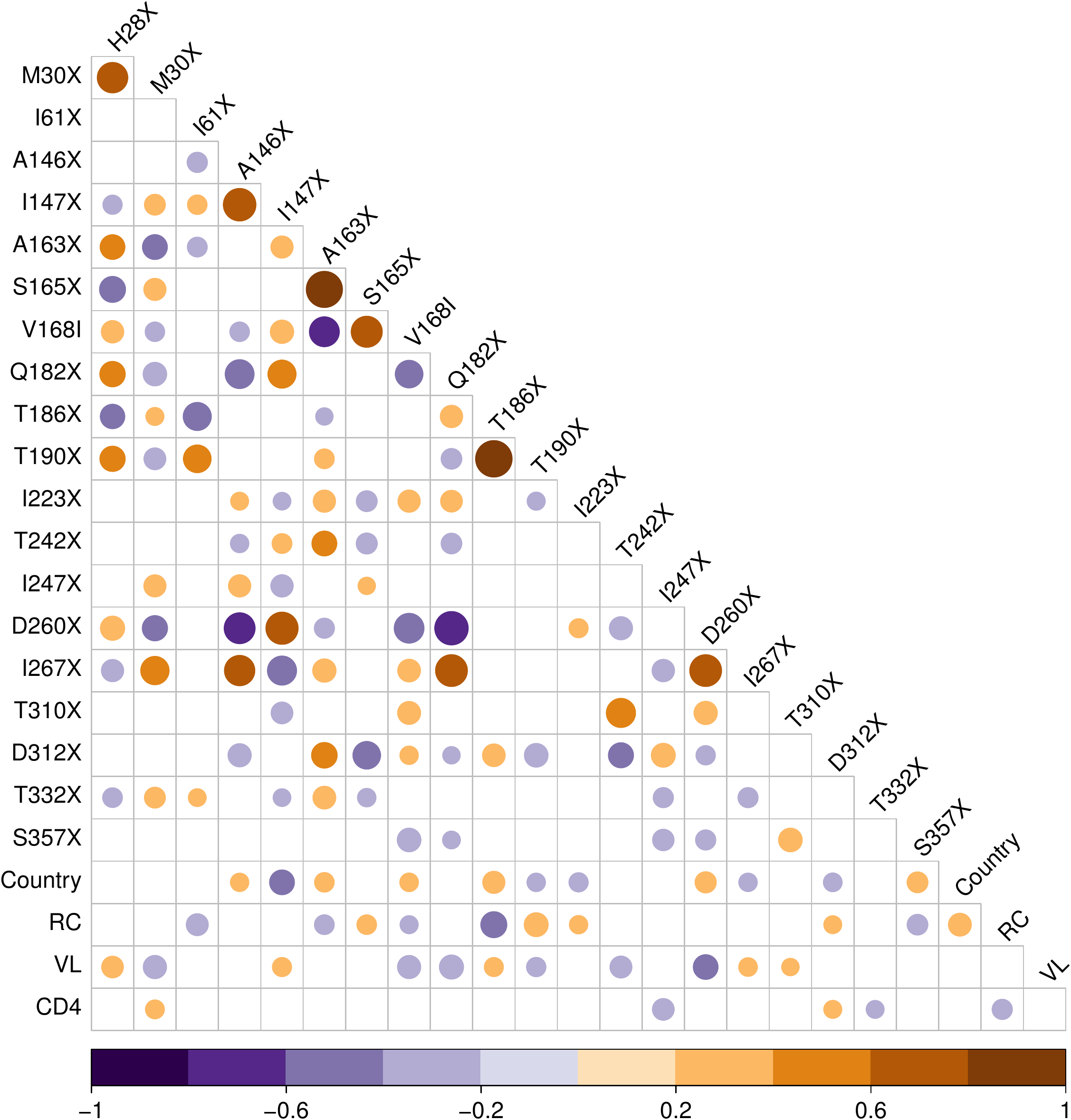}
		\caption{Partial correlation}
		\label{fig:hiv_pcorr}
		\bigskip
	\end{subfigure}
	\begin{subfigure}[t]{0.9\linewidth}
		\centering
		\includegraphics[width=\textwidth]{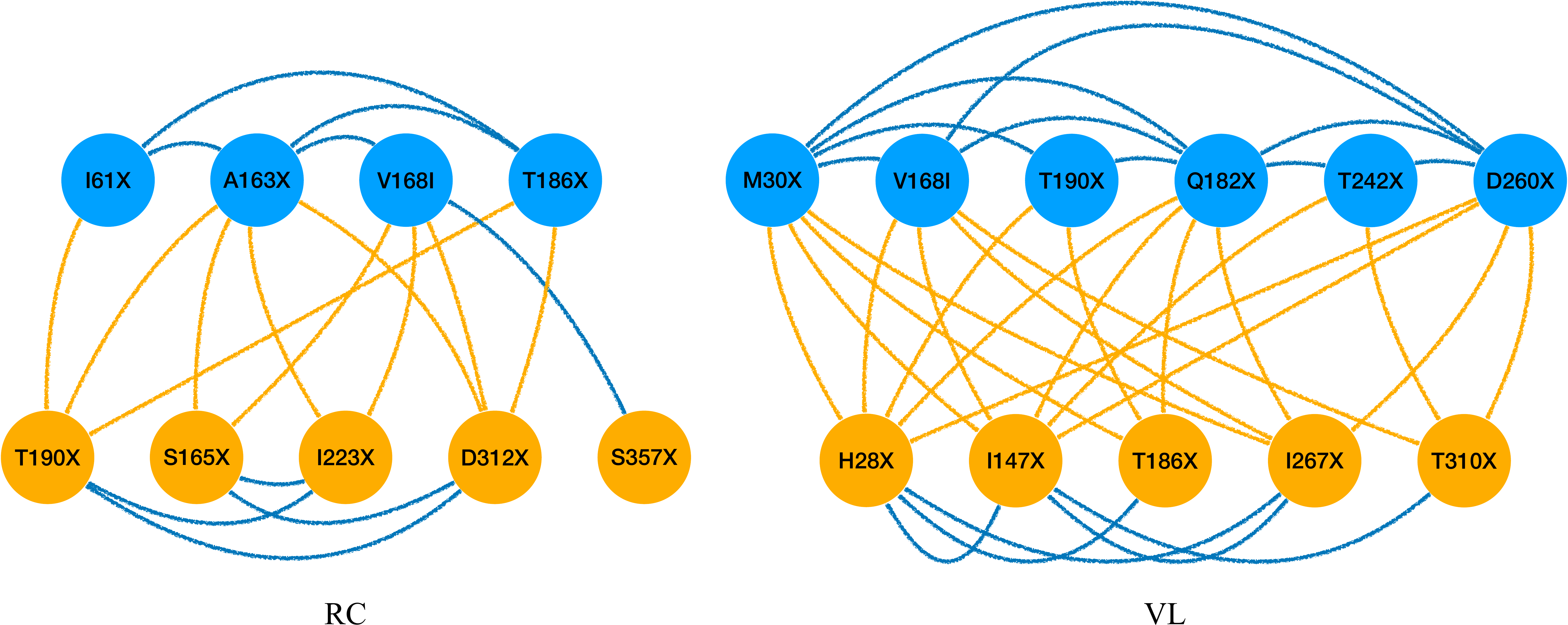}
		\caption{Conditional dependencies among mutations}
		\label{fig:hiv_net}
		\centering
	\end{subfigure}
	\caption{(a) Across-trait correlation and (b) partial correlation with a posterior median $>0.2$ or $< -0.2$ (in color). 
		HIV \textit{gag} mutation names start with the wild type amino acid state, followed by the amino acid site number according to the HXB2 reference genome and end with the amino acid as a result of the mutation (`X' means a deletion).
		Country = sample region: 1 = South Africa, -1 = Botswana; RC = replicative capacity; VL = viral load; CD4 = CD4 cell count.
		(c) Conditional dependencies between HIV-1 immune escape mutations that affect RC or VL.
		Node and edge color indicates whether the dependence is positive (orange) or negative (blue). }
	\label{fig:hiv}
\end{figure}

The data contain $ \nTaxa = 535$ aligned HIV-1 \textit{gag} gene sequences collected from 535 patients between 2003 and 2010 in Botswana and South Africa \citep{Payne2014}.
Each sequence is associated with 3 continuous and 21 binary traits.
The continuous virulence measurements are replicative capacity (RC), viral load (VL) and cluster of differentiation 4 (CD4) cell count.
The binary traits include the existence of HLA-associated escape mutations at 20 different amino acid positions in the \textit{gag} protein and another trait for the sampling country (Botswana or South Africa).
Figure \ref{fig:hiv} depicts across-trait correlations and partial correlations with posterior medians $ > 0.2 $ (or $ < -0.2 $).
Compared to correlations (Figure \ref{fig:hiv_corr}), we observe more partial correlations with greater magnitude (Figure \ref{fig:hiv_pcorr}).
They indicate conditional dependencies among traits after removing effects from other variables in the model, helping to explore the causal pathway.
For example, we only detect a negative conditional dependence between RC and CD4.
In other words, holding one of CD4 and RC as constant, the other does not affect VL, suggesting that RC increases VL via reducing CD4.
The fact that RC is not found to share a strong conditional dependence with VL may be explained by the strong modulatory role of immune system on VL.
Only when viruses with higher RC also lead to more immune damage, as reflected in the CD4 count, higher VL may be observed as a consequence of less suppression of viral replication.
As such, our findings are in line with the demonstration that viral RC impacts HIV-1 immunopathogenesis independent of VL \citep{claiborne2015replicative}.

The partial correlation also helps to decipher epistatic interactions and how the escape mutations and potential compensatory mutations affect HIV-1 virulence.
For example, we find a strong positive partial correlation between T186X and T190X.
Studies have shown that T186X is highly associated with reduced VL \citep{huang2011progression,Wright:2010en} and it requires T190I to partly compensate for this impaired fitness so the virus stays replication competent \citep{wright2012impact}.
The negative conditional dependence between T186X and RC and the positive conditional dependence between T190I and RC are consistent with this experimental observation.
In contrast, with the strong positive association between T186X and T190, the marginal association fails to identify their opposite effects on RC.
Another pair of mutations that potentially shows a similar interaction is H28X and M30X, which have a positive and negative partial correlation with VL, respectively.
These mutations have indeed been observed to co-occur in \textit{gag} epitopes from longitudinally followed-up patients \citep{Olusola2020}.
Figure \ref{fig:hiv_pcorr} keeps all the other compensatory mutation pairs in Figure \ref{fig:hiv_corr} such as A146X-I147X and A163X-S165X that find confirmation in experimental studies \citep{Crawford:2007wn, Troyer:2009ol}.

More generally, when considering the viral trait RC and the infection trait VL, for which their variation are to a considerable extent attributable to viral genetic variation \citep{Blanquart:2017tg}, we reveal an intriguing pattern.
As in Figure \ref{fig:hiv_net}, when two escape mutations impair virulence, and there is a conditional dependence between them, it is always negative.
When two mutations have opposing effects on these virulence traits, the conditional dependence between them (if present) is almost always positive, with one exception of the negative effect between V168I and S357X.
For example, T186X and I61X both have a negative impact on RC and the negative effect between them suggests that their additive, or even potentially synergistic, impact on RC is inhibited.
Moreover, they appear to benefit from a compensatory mutation, T190X, which has been corroborated for the T186X-T190X pair at least as reported above.
Also for VL, the conditional dependence between mutations that both have a negative impact on this virulence trait is consistently negative.
Several of these individual mutations may benefit from H28X as a compensatory mutation, as indicated by the positive effect between pairs that include this mutation, and as suggested above for H28X - M30X.
This illustrates the extent to which escape mutations may have a negative impact on virulence and the need to evolve compensatory mutations to restore it.
We note that our analysis is not designed to recover compensatory mutations at great length as we restrict it to a limited set of known escape mutations, while mutations on many other sites may be compensatory.
In fact, our analysis suggests that some of the considered mutations may be implicated in immune escape due to their compensatory effect rather than a direct escape benefit.

\subsection{Efficiency gain from the new inference scheme}\label{sec:effCompare}
We demonstrate that the joint update of latent variables $\latentData$ and the covariance matrix $\traitCovariance$ significantly improve inference efficiency.
Table \ref{tbSplithmc} compares the performance of four sampling schemes on the HIV immune escape example (described in Section~\ref{sec:hiv}) with $ \nTaxa = 535, \nTraitsDiscrete = 21, \nTraitsCont = 3 $.
We choose our efficiency criterion to be the per run-time ESS for the across-trait correlation $ \traitCorr = \{\traitCorrElement_{ij}\}$ and partial correlation $ \traitPcorr = \{\traitPcorrElement_{ij}\} $ that are of chief scientific interest.
BPS and Zigzag-HMC only update $ \latentData $ and we use the standard NUTS transition kernel (i.e.\ standard HMC combined with no-U-turn algorithm) for the $ \traitCovariance $ elements.
LG-HMC employs the joint update of $ \latentData $ and $ \traitCovariance $ described in Section \ref{sec:jointUpdate}. 
LG-NUTS additionally employs the No-U-Turn algorithm to decide the number of steps and a primal-dual averaging algorithm to calibrate the step size.
We set the same $\totalTime$ for BPS and Zigzag-HMC for a fair comparison.
To tune LG-HMC, we first supply it with an optimal step size $ \stepsize $ learned by LG-NUTS, then decide the number of steps $ \mstep = 100$ as it gives the best performance among the choices (10, 100, 1000). 
As reported in Table \ref{tb:Splithmc}, it is indeed harder to infer partial correlations than correlations and jointly updating $\latentData $ and $ \traitCovariance $ largely eliminates this problem.
BPS loses to the three other samplers and LG-HMC performs the best in terms of ESS for $\traitPcorrElement_{ij}$, yielding a 5$\times$ speed-up.
Without the joint update of  $\latentData $ and $ \traitCovariance $, Zigzag-HMC is only slightly more efficient than BPS.
While a formal theoretical analysis is beyond the scope of this work, we provide an empirical explanation for the different performances of BPS and Zigzag-HMC in Appendix \ref{sec:HZZbeatsBZZ}.
Compared to the manually optimized LG-HMC, LG-NUTS has a slightly lower efficiency likely because the No-U-Turn algorithm requires simulating trajectory both forward and backward to maintain reversibility and this process incurs additional steps \citep{hoffman2014nuts}.
In practice, we recommend using the tuning-free LG-NUTS.
\tbSplithmc

\subsection{Glycosylation of Influenza A virus H1N1}
\label{sec:glyco}
Influenza A viruses of the H1N1 subtype currently circulate in birds, humans, and swine \citep{webster1992evolution, song2008transmission, trovao2020pigs}, where they are responsible for substantial morbidity and mortality \citep{ boni2013economic, ma2020swine}.
The two surface glycoproteins hemagglutinin (HA) and neuraminidase (NA) interact with a cell surface receptor and so their characteristics largely affect virus fitness and transmissibility.
Mutations in the HA and NA, particularly in their immunodominant head domain, sometimes produce glycosylations that shield the antigenic sites against detection by host antibodies and so help the virus evade antibody detection \citep{skehel1984carbohydrate, hebert1997number, daniels2003n, ostbye2020n}.
On the other hand, glycosylation may interfere with the receptor binding and also be targeted by the innate host immunity to neutralize viruses.
Therefore there must be an equilibrium between competing pressures to evade immune detection and maintain virus fitness \citep{tate2014playing, lin2020role}.
The number of glycosylations that leads to this balance is expected to vary in host species experiencing different strengths of immune selection.
Despite decades of tracking IAVs evolution in humans for vaccine strain selection and recent expansions of zoonotic surveillance, the evolvability and selective pressures on the HA and NA have not been rigorously compared across multiple host species.
Here, we examine the conditional dependence between host type and multiple glycosylation sites by estimating the
posterior distribution of across-trait partial correlation while jointly inferring the IAVs evolutionary history.

We use hemagglutinin (H1) and neuraminidase (N1) sequence data sets for influenza A H1N1 produced by Trov\~{a}o et al. as described in \citet{nidia2022}.
We scan all H1 and N1 sequences to identify potential N-linked glycosylation sites, based on the motif Asn-X-Ser/Thr-X, where X is any amino acid other than proline (Pro)  \citep{mellquist1998amino}.
We then set a binary trait for each sequence encoding for the presence or absence of glycosylations at a particular amino acid site.
We keep sites with a glycosylation frequency between 20\% and 80\% for our analysis.
This gives six sites in H1 and four sites in N1.
We include another binary
trait for the host type being mammalian (human or swine) or avian, so the sample sizes are $ \nTaxa = 964, \nTraits = 7 $ (H1) and $ \nTaxa = 896, \nTraits = 5 $ (N1).

\begin{figure}
	\includegraphics[width=\textwidth]{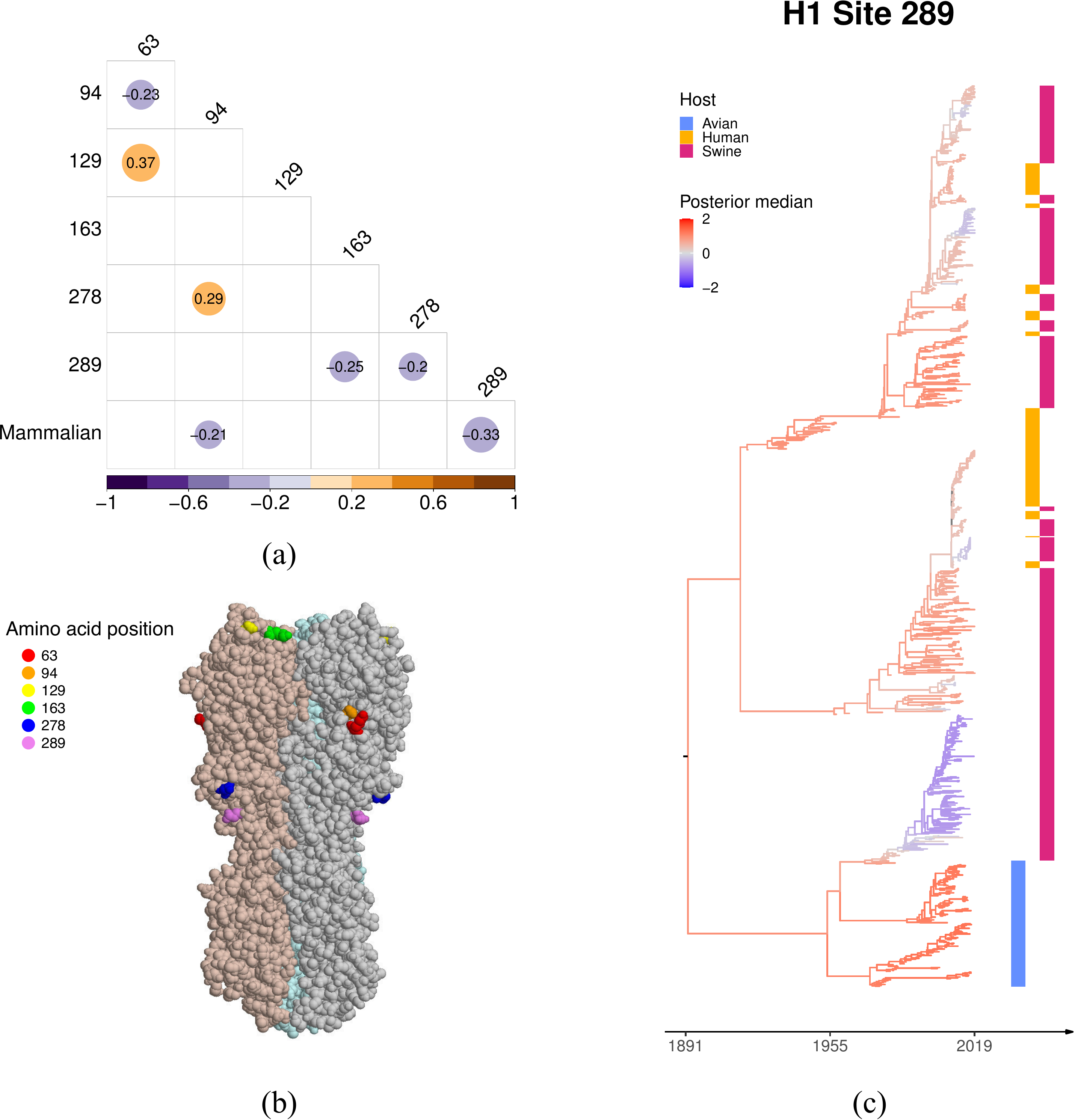}
	\caption
	{%
		(a) Across-trait partial correlation among H1 glycosylation sites and host type with a posterior median $>0.2$ or $< -0.2$ (in color and number).
		(b) HA structure of a 2009 H1N1 influenza virus (PDB entry 3LZG) with six glycosylation sites highlighted.
		Site 278 and 289 are in the stalk domain and all others are in the head domain.
		(c) The maximum clade credibility (MCC) tree with branches colored by the posterior median of the latent variable underlying H1 glycosylation site 289.
		The heatmap on the right indicates the host type of each taxon.
	}%
	\label{fig:H1_merged}
\end{figure}

The six H1 glycosylation sites consist of three pairs that are physically close (63/94, 129/163, and 278/289, see Figure \ref{fig:H1_merged}).
Sites 63 and 94 are particularly close to each other, though distances will vary slightly with sequence.
A negative conditional dependence suggests glycosylation at two close sites may be harmful for the virus (63/94 and 278/289) while a positive effect between two sites suggests a potential benefit (63/129 and 94/278).
We detect a negative conditional dependence between mammalian host and glycosylation site 94 and 289.
Avian viruses have a stronger tendency to have site 289 glycosylated (Figure \ref{fig:H1_merged}).
\begin{figure}[htp]
	\includegraphics[width=\textwidth]{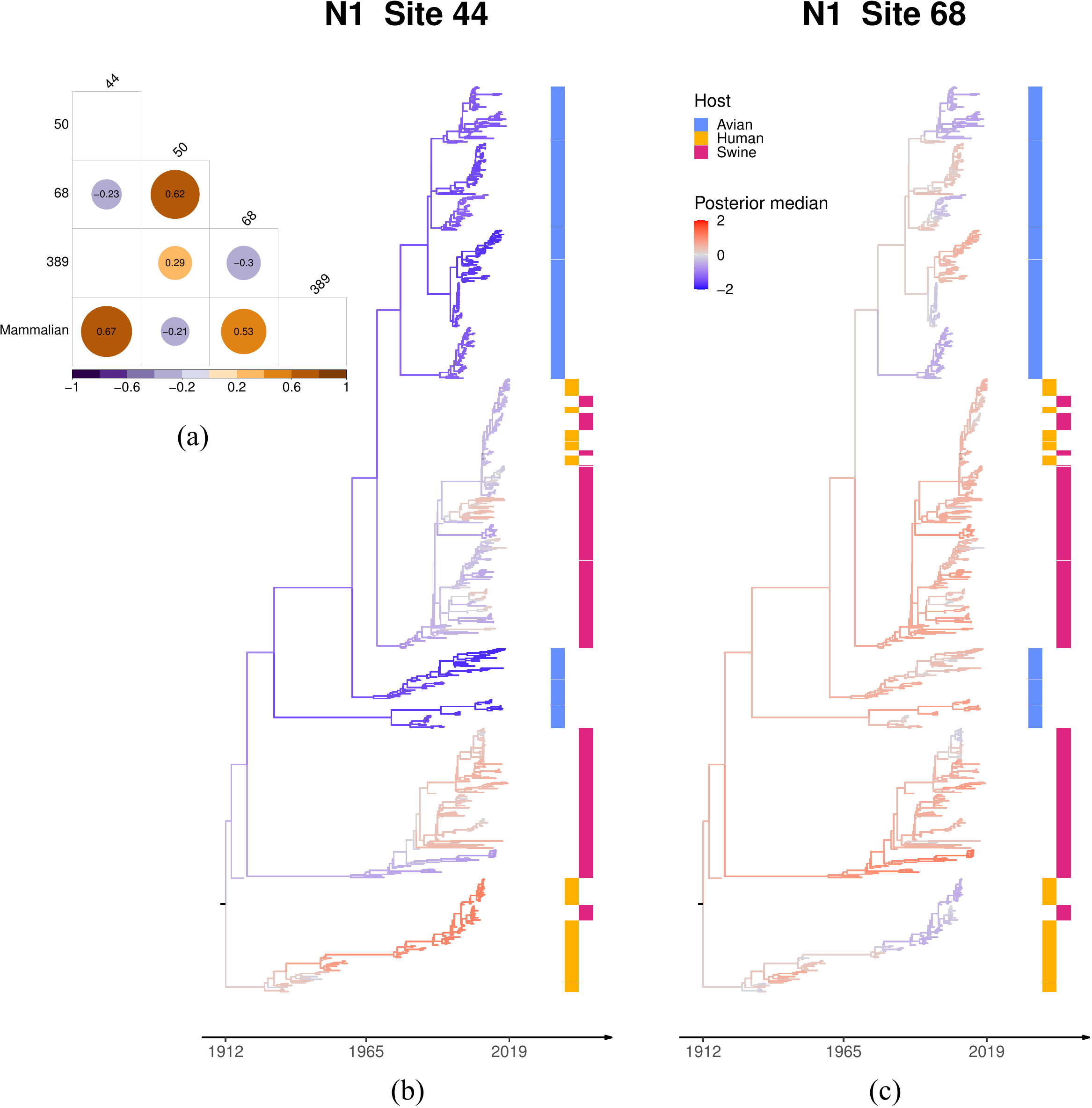}
	\caption{(a) Across-trait partial correlation among N1 glycosylation sites and host type with a posterior median $>0.2$ or $< -0.2$ (in color and number).
	(b)(c) The maximum clade credibility (MCC) tree with branches colored by the posterior median of the latent variable underlying N1 glycosylation site 44 and 68.}
	\label{fig:N1_merged}
\end{figure}
In N1, glycosylations are more strongly correlated than H1 (Figure \ref{fig:N1_merged}).
Two pairs of glycosylation sites have a positive conditional dependency in between (50/68 and 50/389) and two pairs (44/68 and 68/389) have a negative one.
We omit a structural interpretation since all sites but 389 are located in the NA stalk, for which no protein structure is available.
There is a positive conditional dependence between mammalian host and glycosylations at sites 44 and 68.
None of the avian lineages has glycosylation site 44 while most swine and some human lineages have it.
Similarly, glycosylation at site 68 is present in most swine and human lineages but only in avian lineages circulating in wild birds, not those in poultry.

\subsection{\textit{Aquilegia} flower and pollinator co-evolution}
\label{sec:flower}
\begin{figure}[htp]
	\begin{subfigure}[b]{0.5\textwidth}\centering
		\includegraphics[scale=0.3]{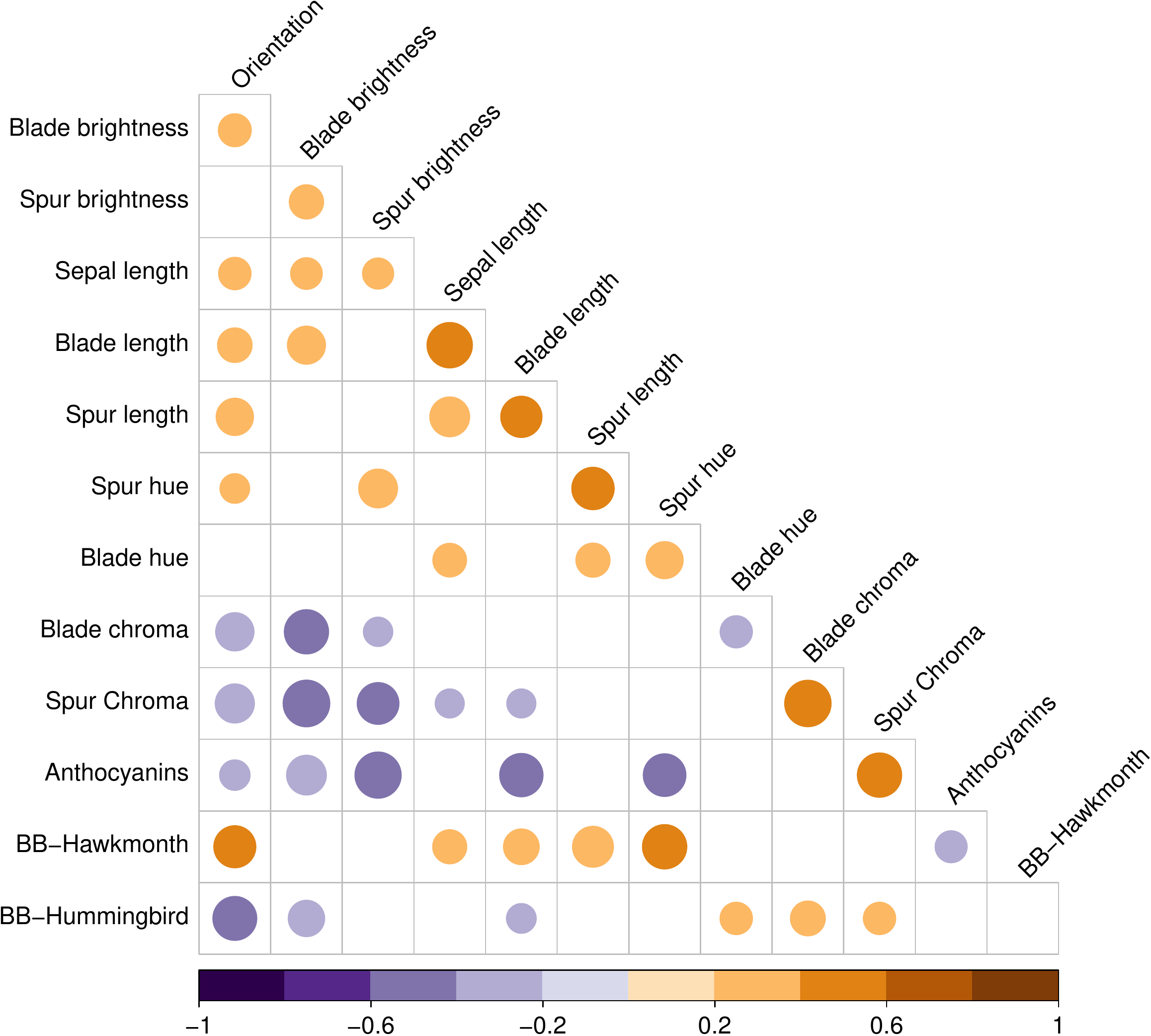}
		\caption{Correlation}
		\label{fig:flower_corr}
	\end{subfigure}
	\begin{subfigure}[b]{0.5\textwidth}\centering
		\includegraphics[scale=0.3]{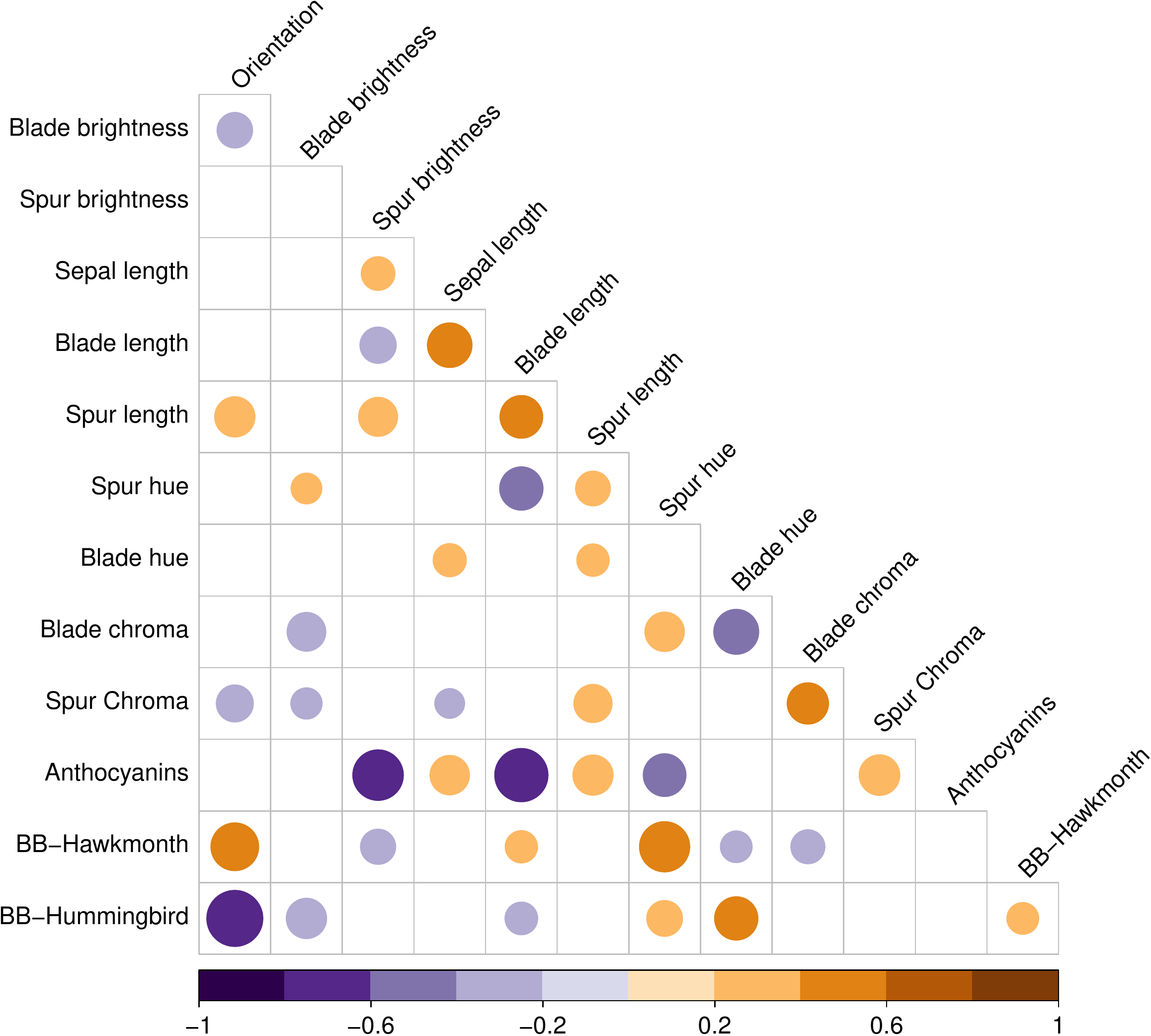}
		\caption{Partial correlation}
		\label{fig:flower_pcorr}
	\end{subfigure}
	\caption{Across-trait correlations and partial correlations with posterior medians $>0.2$ or $< -0.2$ (in color). BB = bumblebee.}
	\label{fig:aquilegia}
\end{figure}

Reproductive isolation allows two groups of organisms to evolve separately, eventually forming new species.
For plants, pollinators play an important role in reproductive isolation \citep{lowry2008strength}.
We examine the relationship between floral phenotypes and the three main pollinators for the columbine genus \textit{Aquilegia}: bumblebees, hummingbirds, and
hawk moths \citep{whittall2007pollinator}.
Here, the pollinator species represents a categorical trait with three classes and we choose bumblebee with the shortest tongue as the reference class.
Figure \ref{fig:aquilegia} provides the across-trait correlation and partial correlation.
Compared to a similar analysis on the same data set that only looks at correlation or marginal association \citep{Cybis2015}, partial correlation controls confounding and indicates the conditional dependencies between pollinators and floral phenotypes that can bring new insights.

For example, we observe a positive marginal association between hawk moth pollinator and spur length but no conditional dependence between them.
The marginal association matches with the observation that flowers with long spur length have pollinators with long tongues \citep{whittall2007pollinator,rosas2014quantitative}.
The absence of a conditional dependence makes intuitive sense because hawk moth's long tongue is not likely to stop them from visiting a flower with short spurs when the other floral traits are held constant.
In fact, researchers observe that shortening the nectar spurs does not affect hawk moth visitation \citep{fulton1999floral}.
Similarly, the positive partial correlation between orientation and hawk moth also finds experimental support.
The orientation trait is the angle of flower axis relative to gravity, in the range of (0, 180).
A small orientation value implies a pendent flower whereas a large value represents a more upright flower \citep{hodges2002genetics}.
Due to their different morphologies, hawk moths prefer upright flowers while hummingbirds tend to visit pendent ones.
Making the naturally pendent \textit{Aquilegia formosa} flowers upright increases hawk moth visitation \citep{hodges2002genetics}.
These results suggest that partial correlation may have predictive power for results from carefully designed experiments with controlled variables.

\subsection{MCMC setup and convergence assessment}\label{sec:converge}
We run all simulations on a node equipped with AMD EPYC 7642 server processors.
For every MCMC run, the minimal effective sample size (ESS) across all dimensions of $ \latentData $ and $ \traitPcorr $ after burn-in is above 100.
As another diagnostic, for our two large-scale applications (Section \ref{sec:glyco} and \ref{sec:hiv}) we run three independent chains and confirm the potential scale reduction statistic $ \hat{R} $ for all partial correlation elements falls between [1, 1.03], below the common criterion of 1.1 \citep{gelman1992inference}.
To reach a minimal ESS = 100 across all $ \traitPcorr $ elements, the post burn-in run-time and number of MCMC transition kernels applied for the joint inference are 21 hours and $ 1.3\times10^6 $ (HIV-1), 113 hours and $ 7.9\times10^7 $ (H1), 76 hours and $ 1.4\times10^8 $ (N1). 
These run-times suggest the difficulty of our large-scale inference tasks where besides the main challenge of sampling $ \{\latentData, \traitCorr, \traitDiag\} $, updating the many tree parameters with Metropolis-Hastings transition kernels also takes a large number of iterations. 
\section{Discussion}
Learning how different biological traits interact with each other from many evolutionarily related taxa is a long-standing problem of scientific interest that sheds light on various aspects of evolution.
Towards this goal, we develop a scalable solution that significantly improves inferential efficiency compared to established state-of-the-art approaches \citep{Cybis2015,zhang2021large}.
Our novel strategy enables learning across-trait conditional dependencies that are more informative than the previous marginal association based analyses.
This approach provides reliable estimates of across-trait partial correlations for large problems, on which the established BPS-based method struggles.
In two large-scale analyses featuring HIV-1 and H1N1 influenza, the improved efficiency allows us to infer conditional dependencies among traits of scientific interest and therefore investigate some of the most important molecular mechanisms underlying the disease.
In addition, our approach incorporates automatic tuning, so that the most influential tuning parameters automatically adapt to the specific challenge the target distribution presents.
Finally, we extend the phylogenetic probit model to include categorical traits and illustrate its use in examining the co-evolution of \textit{Aquilegia} flower and pollinators.

We leverage the cutting-edge Zigzag-HMC \citep{nishimura2020discontinuous} to tackle the exceedingly difficult computational task of sampling from a high-dimensional truncated normal distribution in the context of the phylogenetic probit model.
Zigzag-HMC proves to be more efficient than the previously optimal approach that uses the BPS (Section \ref{sec:effCompare}), especially when combined with differential operator splitting to jointly update two sets of parameters $ \latentData $ and $ \traitCovariance $ that are highly correlated. 
The improved efficiency allows us to obtain reliable estimates of the conditional dependencies among traits.
In our applications, we find that these conditional dependencies better describe trait interactions than do the marginal associations.
It is worth mentioning that another closely related sampler, the Markovian zigzag sampler \citep{bierkens2019zig}, or MZZ, may also be appropriate for this task but provides lower efficiency than Zigzag-HMC \citep{nishimura2021hamiltonian}.
While Zigzag-HMC is a recent and less explored version of HMC, BPS and MZZ are two central  methods within the piecewise deterministic Markov process literature that have attracted growing interest in recent years \citep{fearnhead2018piecewise, dunson2020hastings}.
Intriguingly, the most expensive step of all three samplers is to obtain the log-density gradient, and the same linear-order gradient evaluation method \citep{zhang2021large} largely speeds it up.

We now consider limitations of this work and the future directions to which they point.
First, the phylogenetic probit model does not currently accommodate a directional effect among traits since it only describes pairwise and symmetric correlations.
However, the real biological processes are often not symmetric but directional, where it is common that one reaction may trigger another but not the opposite way.
A model allowing directed paths is preferable since it better describes the complicated causal network among multiple traits.
Graphical models with directed edges  \citep{lauritzen1996graphical} are commonly used to learn molecular pathways \citep{neapolitan2014modeling, benedetti2017network}, but challenges remain to integrate these methods with a large and randomly distributed phylogenetic tree.
Toward this goal, one may construct a continuous-time Markov chain to describe how discrete traits evolve \citep{pagel1994detecting, o2012evolutionary}, but with $ \nTraits $ binary traits the transition rate matrix grows to the astronomical size $ 2^\nTraits $.
Second, though our method achieves the current best inference efficiency under the phylogenetic probit model, there is still room for improvement.
In the influenza glycosylation example, we use a binary trait indicating the host being either avian or mammal (human or swine), instead of setting a categorical trait for host type.
In fact, we choose not to use a three-class host type trait because it causes poor mixing for the partial correlation elements.
We suspect two potential reasons for this.
First, according to our model assumptions for categorical traits (Equation \refeq{eq:thresholdFunc}), the latent variables underneath the same trait are very negatively correlated, leading to a more correlated and challenging posterior.
Second, in our specific data sets, the glycosylation sites tend to be similar in human and swine viruses, further increasing the correlation among posterior dimensions.
One potential solution is to de-correlate some latent variables by grouping them into independent factors using phylogenetic factor analysis \citep{tolkoff2018phylogenetic, hassler2021principled}.
Finally, one may consider a logistic or softmax function to map latent variables to the probablity of a discrete trait.
This avoids the hard truncations in the probit model but also adds another layer of noise.
It requires substantial effort to develop an approach that overcomes the above limitations while supporting efficient inference at the scale of applications in this work.

\section{Acknowledgments}
We thank Kristel Van Laethem for useful discussion about HIV replicative capacity, CD4 counts and viral load.
ZZ, PL and MAS are partially supported by National Institutes of Health grant R01 AI153044.
MAS and PL acknowledge support from the European Research Council under the European Union's Horizon 2020 research and innovation programme (grant agreement no.~725422 - ReservoirDOCS) and from the Wellcome Trust through project 206298/Z/17/Z (The Artic Network).
JLC is supported by the intramural research program of the National Library of Medicine, National Institutes of Health.
AH is supported by NIH grant K25AI153816. 
This work uses computational and storage services provided by the Hoffman2 Shared Cluster through the UCLA Institute for Digital Research and Education's Research Technology Group.
The opinions expressed in this article are those of the authors and do not reflect the view of the National Institutes of Health, the Department of Health and Human Services, or the United States government.
\bigskip
\begin{center}
{\large\bf SUPPLEMENTARY MATERIAL}
\end{center}
We implement our algorithms within \textsc{BEAST} \citep{beast2018} and provide the data sets and instructions at \url{https://github.com/suchard-group/hzz_data_supplementary}.

\appendix
\section{Auto-tuning of $ \rs $}\label{sec:tuning}
We describe a simple heuristic to auto-tune the step size ratio $ \rs $ on the fly.
Let $ \grandVariance_G $ and $ \grandVariance_L $ be the covariance matrices for $\bposition_G$ and $ \bposition_L $ respectively, then their minimal eigenvalues $ \lambda_{\text{min}, G}$ and $ \lambda_{\text{min}, L}$ describe the variance magnitude in the most constrained direction.
Intuitively, for both HMC and Zigzag-HMC, the step size should be proportional to the diameter of this most constrained density region, which is $ \sqrt{\lambda_{\text{min}, G}} $ or $ \sqrt{\lambda_{\text{min}, L}} $.
Therefore we propose a choice of $ \rs =  \frac{\sqrt{\lambda_{\text{min}, L}}}{\sqrt{\lambda_{\text{min}, G}}} $, assuming the two types of momenta lead to similar travel distance during one unit time.
It is straightforward to check this assumption.
At stationarity, HMC has a velocity $ \bv_G \sim \normalDistribution{\mathbf{0}}{\mathbf{I}}$, so its velocity along any unit vector $ \bm{u} $ would be distributed as $ \langle\bv_G, \bm{u}\rangle \sim \normalDistribution{0}{1}$, and the travel distance $\mathbb{E}|\langle \bv_G, \bu \rangle| = \sqrt{2 / \pi}$.
For Zigzag-HMC,  as $ \langle\bv_L, \bm{u}\rangle  $ does not follow a simple distribution, we estimate $ \mathbb{E}|\langle \bv_L, \bu \rangle|  $ by Monte Carlo simulation and it turns out to be $ \approx 0.8 $, close to $ \sqrt{2 / \pi} $.

We test this intuitive choice of $ \rs $ on a subset of the HIV data in \citet{zhang2021large} with 535 taxa, 5 binary and 3 continuous traits.
We calculate the optimal  $ \rs = \frac{\sqrt{\lambda_{\text{min}, L}}}{\sqrt{\lambda_{\text{min}, G}}} \approx 2.5$ with $ \grandVariance_G $ and $ \grandVariance_L $ estimated from the MCMC samples.
Clearly, $ \rs $ has a significant impact on the efficiency as a very small or large $\rs$ leads to lower ESS (Table \ref{tbCompareRS}).
Also, an $ \rs $ in the order of our optimal value generates the best result, so we recommend this on-the-fly automatic tuning $ \rs =  \frac{\sqrt{\lambda_{\text{min}, L}}}{\sqrt{\lambda_{\text{min}, G}}}$ (Table \ref{tb:CompareRS}).
\tbCompareRS

\section{Zigzag-HMC explores the energy space more efficiently than BPS}\label{sec:HZZbeatsBZZ}

In our experience, BPS tends to generate samples with high auto-correlation between their respective energy function evaluations $-\logd$.
In other words, it slowly traverses the target distribution's energy contours even when the marginal dimensions all appear to demonstrate good mixing.
A similar behavior has also been reported by \cite{bouchard2018bouncy}, who introduce a velocity refreshment to address the issue.
As we demonstrate below, however, even velocity refreshments cannot fully remedy BPS's slow-mixing on the energy space.

We apply BPS and Zigzag-HMC to a 256-dimensional standard normal truncated to the positive orthant (all $ x_i > 0 $). 
We run both samplers for $2000$ iterations where per-iteration travel time is one unit time interval and repeat the experiments for 10 times with varying initial values.
For BPS we include Poisson velocity refreshments to avoid reducible behavior and set the refreshment rate to an optimal value 1.4 \citep{bierkens2018high}.
At every iteration we refresh Zigzag-HMC's momentum by redrawing it from the marginal Laplace distribution.
Both samplers have no problem sampling from the target distribution and the minimal ESS across all dimensions are $ 158\pm25$ (mean $\pm$ SD) for BPS and $ 207\pm21$ for Zigzag-HMC, estimated from the last 1000 samples of the MCMC chains across 10 runs.
As a sanity check, the average sample mean and variance are $ (0.800, 0.365) $ for BPS and $ (0.798, 0.363) $ for Zigzag-HMC, close to the analytical values --- the univariate marginal distribution of our truncated standard normal is a truncated normal with mean $2/\sqrt{2\pi} \approx 0.798$ and variance $1 - 2/\pi \approx 0.363$ \citep{cartinhour1990one}.

\begin{figure}[h]
	\centering
	\includegraphics[width=0.9\linewidth]{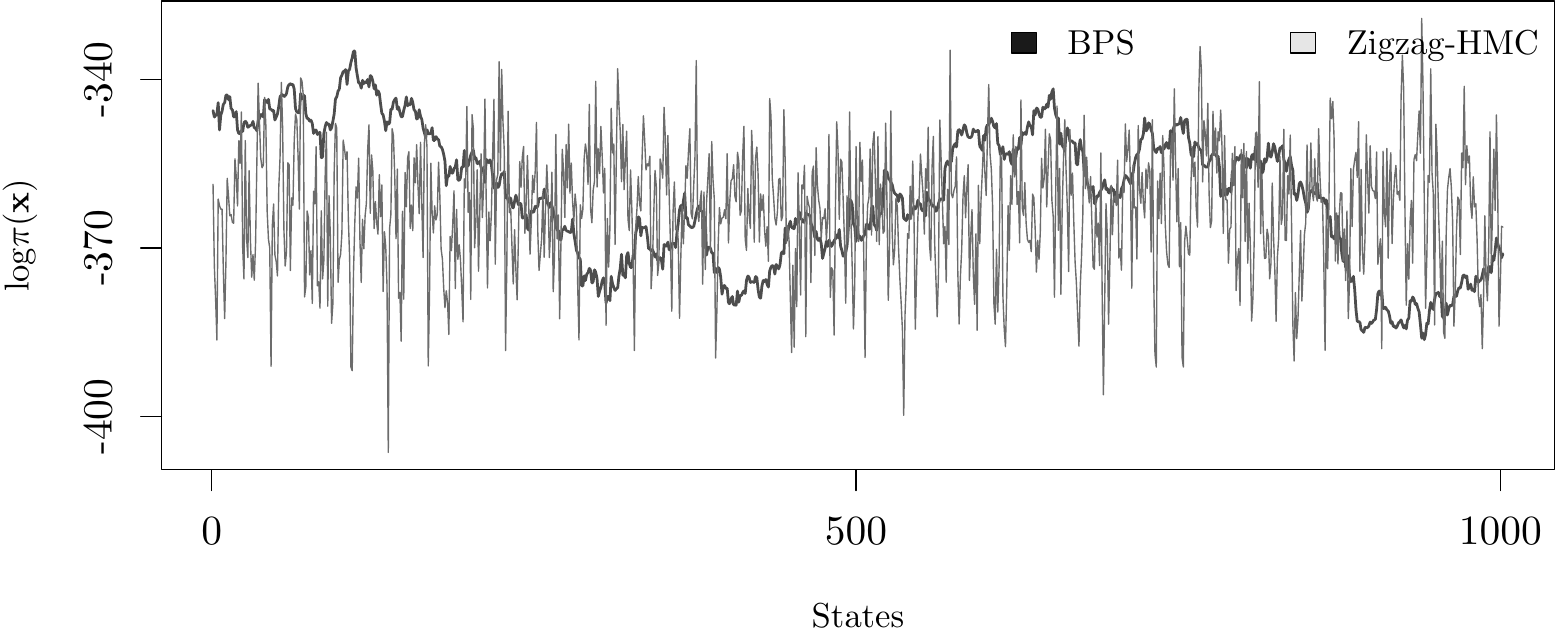}
	\caption{Trace plot of the log density of a $ 256$-dimensional truncated standard normal sampled by BPS and Zigzag-HMC for 1000 MCMC iterations.}
	\label{fig:traceJD}
\end{figure}

However, Zigzag-HMC returns a clear win over BPS in the mixing of joint density (Figure \ref{fig:traceJD}). 
The sampling inefficiency for $ -\logd $  is less of a problem if one only needs to sample from a truncated normal with a fixed covariance matrix, but we are keenly interested in sampling the covariance matrix as a target of scientific interest. In this context, inefficient traversal across energy contours harms the sampling efficiency for all model parameters (Section \ref{sec:effCompare}).

We can provide an intuition for BPS's slow movement in energy space.
Assume the $ d $-dimensional parameter at the $ \seqT $th MCMC iteration is $ \parameterVec(\seqT) = \left(\parameter{1}(\seqT), \dots, \parameter{d}(\seqT) \right)\in \realNumbers^d$, $ \seqT = 1, \dots, \seqLength $, with $ \seqLength $ being the total number of iterations.
For a truncated standard normal, its log density $\logd \propto \sum_{i}^{d} \parameter{i}^2$, and a high auto-correlation suggests $ \logd $ changes little between successive iterations, that is, the squared jumping distances
\begin{equation*}
\jumpDis = \left[\sum_{i}^{d} \parameter{i}^2(t+1) -\sum_{i}^{d} \parameter{i}^2(t) \right]^2, \quad   t = 0, \dots, \seqLength-1
\end{equation*}
\vspace{-2mm}
are small. We then decompose $ \jumpDis $ into two components
\begin{equation}
\begin{aligned}
\jumpDis &= \termOne + \termTwo, \\
\termOne &= \sum_{i}^{d} \left[\parameter{i}^2(t+1) - \parameter{i}^2(t)\right]^2 , \\
\termTwo &= \sum_{j\neq k}^{d} \left[\parameter{j}^2(t+1) - \parameter{j}^2(t)\right] \left[\parameter{k}^2(t+1) - \parameter{k}^2(t)\right],\: t = 0, \dots, \seqLength-1,
\end{aligned}
\end{equation}
where $\termOne$ measures the sum of the marginal travel distances and $ \termTwo$ the covariance among them.
We compare $ \jumpDis$, $\termOne$ and $\termTwo $ between BPS and Zigzag-HMC in the aforementioned experiments.
Clearly seen in Table \ref{tb:JDanalysis}, BPS yields a much lower $ \jumpDis $ than Zigzag-HMC because its $ \termTwo $ is largely negative, suggesting strong negative correlation among the coordinates.
\tbJDanalysis
\clearpage
\bibliographystyle{chicago}
\bibliography{hzz_app}
\end{document}